**Title:**

Co-tunneling spin blockade observed in a three-terminal triple quantum dot

**Authors:**

A. Noiri[1,2], T. Takakura[1], T. Obata[1], T. Otsuka[1,2,3], T. Nakajima[1,2], J. Yoneda[1,2], and S. Tarucha[1,2]

**Affiliations:**

[1]*Department of Applied Physics, University of Tokyo, 7-3-1 Hongo, Bunkyo-ku, Tokyo 113-8656, Japan*

[2]*RIKEN, Center for Emergent Matter Science (CEMS), Wako-shi, Saitama 351-0198, Japan*

[3]*JST, PRESTO, 4-1-8 Honcho, Kawaguchi, Saitama, 332-0012, Japan*

**Abstract:**

We prepare a triple quantum dot with a separate contact lead to each dot to study Pauli spin blockade in the tunnel-coupled three dots in a row. We measure the tunneling current flowing between the center dot and either the left or right dot with the left and right leads as a common source and the center lead as a drain. In the biased stability diagram, we establish Pauli spin blockade in the respective neighboring dots, with features similarly obtained in double quantum dot systems. We further realize Pauli spin blockade with two different conditions by tuning the inter-dot coupling gates: strong and weak inter-dot tunnel coupling regimes. In the strong-coupling regime we observe significant suppression of co-tunneling through the respective double dots due to Pauli spin blockade. We reveal the influence from the third dot in the triple dot device on this co-tunneling Pauli spin blockade and clarify that the co-tunneling Pauli spin blockade is lifted by the resonant coupling of excited states to the third dot level as well as spin exchange of the left and right dots with the adjacent reservoir.

**Introduction:**

Coulomb blockade (CB) and Pauli spin blockade (PSB) are both core concepts to represent the interaction effects of electrons confined to quantum dots (QDs), and have been studied to explore the underlying physics [1] and also to implement various techniques of spin-based quantum computing [2-4]. CB arises from the Coulombic interaction within QDs [5] and between QDs [6]. The charge states in QDs are precisely defined by CB and well resolved in the charge stability diagram [6], which is obtained using either tunneling current measurement or charge sensing [7]. CB is partially lifted by co-tunneling [8] through the intermediated states when the temperature is elevated or a source-drain bias is applied.

On the other hand, PSB is associated with specific spin states and is typically observed in double quantum dots (DQDs) [9-11]: when the two-electron state is an excited triplet state having one electron in each dot, labeled (1,1), the transition to the ground singlet state having two electrons in one of the two dots, labeled (0,2), is, although energetically allowed, forbidden due to Pauli exclusion. PSB is lifted upon spin flip or spin relaxation due to spin-orbit interctions or fluctuating nuclear field. Especially the latter is predominant at a low external field in semiconductor QDs such as GaAs and InAs QDs, which contain a nuclear spin bath, and the PSB lifting appears when the blockaded (1,1) triplet is mixed with the singlet of (1,1). PSB can be identified in the biased stability diagram, which is usually derived by measuring transport current through DQDs with a finite source-drain voltage [9-11]. Lifting of PSB accompanies a charge move from (1,1) to (0,2), and this is detected using a charge sensor or measurement of an excess transport current flow subsequent to the charge move. Note CB is more robust than PSB and its lifting due to co-tunneling is observed as an excess transport current along the boundaries between charge states in the biased stability diagram [6]. The co-tunneling is usually treated for spinless electrons, but it should depend on spin configuration when the transition to the intermediate state is spin-blocked, although it has not been addressed to date. We call this co-tunneling spin blockade Co-PSB. The co-tunneling we study here is inelastic because the energy of the electron state before and after the co-tunneling is different [8].

Studies on CB and PSB are now moving onto multiple QDs (MQDs) such as triple [12-16], quadruple, and quintuple QDs [17-20], where the charge states become subtler to distinguish and more complicated to control. In addition, the transport current through series of dots is strongly reduced except for the resonance points [12,14,21]. Therefore, measurement of the biased stability diagram of MQDs including CB, PSB and their lifting due to co-tunneling needs technical ingenuity. In these MQDs PSB is usually treated for neighboring pairs of QDs in the same way as for DQDs, but it can be different if it is influenced by coupling of the PSB states to the states in the remaining dot or dots [22].

In this work we use a three-terminal triple QD (3T-TQD) having a tunnel-coupled lead to each dot [22] to study CB and PSB with influence from the third dot. With this device, we are able to observe CB and PSB in TQD by measuring transport current between the center dot and the right or left dot. A similar type of 3T-TQD with a relatively weak inter-dot tunnel coupling was previously used to study cooperative lifting of PSB [22]. On the other hand, we use 3T-TQDs with a relatively strong inter-dot tunneling to study PSB in the left two dots (L-DQD), or right two dots (R-DQD) being influenced by the rightmost dot, or leftmost dot, respectively. We find that the empty third dot can influence Co-PSB in the adjacent DQD because Co-PSB in either DQD is partially lifted due to coupling to the remaining dot. We observe lifting of Co-PSB when the PSB state in L- or R-DQD is resonated with the state in the remaining dot. The present observation indicates that PSB is important in the co-tunneling process as well.

**Results:**

Our device, shown in Fig. 1(a), has Ti/Au gate electrodes on the surface of GaAs/n-AlGaAs substrate to define a TQD in a 100 nm deep two-dimensional electron gas. The center reservoir ($D_C$) is used as a drain and two side reservoirs ($S_L$, $S_R$) as two sources. Two quantum point contacts (QPCs) which are used as charge sensors for the TQD are also defined by the gate electrodes. The TQD device is placed in a dilution refrigerator with a base temperature of ~50 mK. We measure the sum of the two dot currents $I_{QD}$ flowing from $D_C$ to $S_L$ for L-DQD and from $D_C$ to $S_R$ for R-DQD. In this measurement we apply a finite source-drain bias, $V_{SD}$ to $D_C$ and ground all of the other Ohmic contacts. We also measure the sum of the two charge sensor currents $I_{QPC}$ to probe the charge states by applying a finite voltage, $V_{QPC}$ to $D_L$ and $D_R$ and ground all of the other Ohmic contacts.

The unbiased, and biased charge stability diagram as a function of two gate voltages $V_{CL}$ and $V_{CR}$ are obtained from measurement of $I_{QPC}$, and $I_{QD}$, respectively, and shown in Fig. 1(b) and (c). Each charge state is well resolved in Fig. 1(b) and the electron occupation ($N_L,N_C,N_R$) in the left, center and right dot, respectively is precisely counted, starting from (0,0,0). Formation of TQD having a few electrons is confirmed from the observation of three families of charging lines with different slopes: large, intermediate, and small for the charging of the left, center, and right dot, respectively. In this work we assume a 3T-TQD as a combined system of L-DQD (or R-DQD) having two electrons and the remaining dot having one or no electron. Therefore, for convenience we use notations of two-electron spin states in DQD to specify the spin states of 3T-TQD, such as a singly occupied state of $(1,1,0)S_{LC}$ $((0,1,1)S_{CR})$ and a doubly occupied state in the center dot of $(0,2,0)S_C$ for the singlet states and $(1,1,0)T_{\alpha,LC}$ $((0,1,1)T_{\alpha,CR})$ and $(0,2,0)T_{\alpha,C}$ for the triplet states of L-DQD (and similarly for R-DQD). Here, $T_\alpha$ is a triplet state having the z-component of +1, 0, and -1 for $T_+$, $T_0$, and $T_-$, respectively. First, we show PSB in both of L-DQD and R-DQD with strong and weak inter-dot tunnel couplings to demonstrate the tunability of our device. We find the observed PSB features in these measurements agrees with those of PSB in DQDs [10]. Then we focus on the strong coupling condition to study the co-tunneling effect in PSB with influence from the third dot.

When we apply a source-drain bias of $V_{SD}$ = -0.5 mV, so-called bias triangles [6] are observed in the biased stability diagram near the boundaries of (1,1,0)-(0,2,0) for L-DQD and (0,1,1)-(0,2,0) for R-DQD as shown in Fig. 1(c). The charge states are determined by the charge sensing experiment performed with the same gate voltage condition (the corresponding gate voltage region is shown by the red dashed square in Fig. 1(b)). In L- or R-DQD, $I_{QD}$ flows through the two electron states when the Fermi energy of the source, the electro-chemical potential of (1,1,0) or (0,1,1), the electro-chemical potential of (0,2,0) and the Fermi energy of the drain are aligned in descending (ascending) order with the positive (negative) source-drain bias. This condition is met inside a triangle (so-called bias triangle) which is formed at a crossing of two different charging lines (see Fig. 3(c)). The size of the bias triangle expands in proportion to $|V_{SD}|$ [6] and the direction depends on the bias polarity [6]. Here the effect of PSB appears as $I_{QD}$ suppression in the bias triangle located between the resonance lines of singly-occupied and doubly-occupied singlet states (singlet resonance line) and those of triplet states (triplet resonance line) [9-11]. Similar bias triangles with bias polarity dependence are also observed in the 3T-TQD studied here as in Fig. 2, which shows the zoomed-in data of $I_{QD}$ as a function of $V_{CL}$ and $V_{CR}$ with $B_{ext} = 0$ T measured for $V_{SD} = -1.0$ mV in (a) and $V_{SD} = 1.0$ mV in (b). In (a) a finite $I_{QD}$ always flows inside the bias triangle, while in (b) the size of the bias triangle is the same as in (a) but $I_{QD}$ is strongly suppressed between the singlet and triplet resonance lines shown by the yellow dashed trapezoidal regions due to PSB [10]. Figs. 2(c), and (d) ((e), and (f)) show the energy diagrams on the singlet resonance line at the points marked by the red square, and circle, respectively in Fig. 2(a) (Fig. 2(b)). For the negative bias in (a) an electron enters the center dot from the center reservoir to establish $(0,2,0)S_C$ and moves to either the right or left dot before escaping to the adjacent reservoir. On the other hand, for the positive bias in (b), when an electron enters either the right or left dot from the adjacent reservoir, the charge configuration becomes (1,1,0) or (0,1,1). These two-electron states are either spin singlet or triplet. In the case of triplet, the electron can no longer move to the center dot due to PSB.

In GaAs DQDs, one expects a difference of $\Delta B_n \sim 5$ mT [1,10] in the Overhauser field between QDs due to the statistical fluctuation of nuclear spin bath. This induces mixing of the singlet and triplet states and lifts PSB when the singlet-triplet energy difference, $\Delta E_{ST}$, is smaller than the energy corresponding to $\Delta B_n$ as given by $\Delta E_n = |g|\mu_B \Delta B_n$ (~100 neV). Near the resonance of singly- and doubly-occupied singlet states (the red square and circle in Fig. 2(a) and (b)), $\Delta E_{ST} \approx E_t$ at $B_{ext} = 0$ T [10]. Then PSB behaviors can be classified into two different regimes depending on the strength of the inter-dot tunnel coupling $E_t$: strong ($E_t \gg \Delta E_n$) and weak ($E_t \lesssim \Delta E_n$) inter-dot tunnel coupling [10]. Under the strong inter-dot coupling condition that $\Delta E_{ST} \sim E_t \gg \Delta E_n$, PSB always occurs at $B_{ext} = 0$ T. By applying a finite $B_{ext}$ such that Zeeman energy compensates for $E_t$, $\Delta E_{ST} \sim |\Delta E_t - E_Z| \lesssim \Delta E_n$, PSB is lifted because of the singlet-triplet mixing due to the nuclear spin bath fluctuation [10,23]. Fig. 2(g) shows the $B_{ext}$ dependence of $I_{QD}$ measured near the singlet resonance line (red circle in Fig. 2(b)). $I_{QD}$ is initially low for $B_{ext} < 0.5$ T due to PSB and gradually increases to make a peak at $B_{ext} \approx 1.3$ T where $E_Z \approx E_t$. From this peak position, $E_t$ is roughly estimated as a few tens of μeV ($\gg \Delta E_n$), indicating a strong inter-dot coupling of TQD. A similar magnetic field dependence was previously observed in a DQD having a strong inter-dot tunnel coupling [10]. This peak appears wider than expected from the typical value of $\Delta B_n$ (~5 mT [1,10]) probably because of the effect of dynamic nuclear polarization [10].

To demonstrate the tunabiliy of inter-dot tunnel couplings, we make $V_{CL}$ and $V_{CR}$ more negative and the three plunger gate voltages more positive. Then, we could reduce the inter-dot tunnel coupling sufficiently to satisfy the weak-coupling condition of $E_t \lesssim \Delta E_n$ while keeping the few-electron regime. Inset of Fig. 2(g) shows the $B_{ext}$ dependence of $I_{QD}$ at the singlet resonance line in such a condition. Compared to the strong-coupling case, no current suppression is observed at $B_{ext} = 0$ T because of the influence of the inhomogeneous nuclear field. $I_{QD}$ readily decreases with increasing $B_{ext}$ to $\approx 10$ mT where $\Delta E_{ST} \gtrsim \Delta E_n$ is satisfied. This is a typical behavior observed for the weak-coupling PSB in DQDs [10] ($E_t \lesssim \Delta E_n \sim 100$ neV).

These results described above are consistent with the PSB features previously observed in DQDs [10], but we find that the PSB can be influenced by the remaining dot in certain conditions. This is particularly the case for the second-order tunneling or co-tunneling, which is the major topic discussed in this work. Hereafter we use the strong inter-dot tunnel coupling condition to study the co-tunneling effect on PSB in TQD. First we focus on the CB region of the two-electron state. Fig. 3 shows the biased stability diagram of $I_{QD}$ v.s. $V_{CL}$ and $V_{CR}$ with $V_{SD} = 1.0$ mV measured for $B_{ext} = 0$ T in (a) and $B_{ext} = 1$ T in (b). The CB regions of (1,1,0), (0,1,1) and (1,1,1) and two bias triangles at the boundaries of (1,1,1)-(0,2,1) and (1,1,1)-(1,2,0) are clearly observed. These features are shown schematically in Fig. 3(c). Here we find a couple of interesting features marked by color symbols reflecting the influence on CB and PSB in L- or R-DQD from the tunnel coupling to the remaining dot (red and yellow) as well as spin exchange between the left or right quantum dot and adjacent lead (green). Hereafter we interpret these characters one-by-one.

First, we focus on the feature at the green symbols in Fig. 3(a). $I_{QD}$ in the CB regions of (1,1,0) and (0,1,1) is strongly suppressed except near the charge boundaries of the leftmost and rightmost dots in Fig. 3(a). The current flow along these charging lines is due to co-tunneling through the doubly occupied center dot state $(0,2,0)S_C$. Inside the CB regions this co-tunneling is spin-blocked once $(0,1,1)T_{a,CR}$ is formed in R-DQD as schematically shown in Fig. 3(d) and a similar argument holds for L-DQD in (e). This Co-PSB is lifted along the two-electron charging lines shown by the green square and triangle in Fig. 3(a) due to the spin exchange with the adjacent reservoir (see Fig. 3(f) showing the corresponding energy diagram at the green square in Fig. 3(a)). More or less the same scenario follows for the border between (0,1,1) and (1,1,1) and between (1,1,0) and (1,1,1). Note co-tunneling in the CB region can occur through the doubly occupied triplet state $(0,2,0)T_{a,C}$ in a spin-preserving process but only slowly because the $(0,2,0)T_{a,C}$ state is energetically high. In addition, we find a significantly large current flow at the degenerate point of (0,1,1) and (1,1,0) (shown by the green circle) in Fig. 3(a). The two triple points at two ends of the (0,1,1)-(1,1,0) degenerate line are closely spaced in energy because of the weak electro-static coupling between the end dots of TQD. Therefore co-tunneling through the two-electron states and three-electron states can both contribute to generate the current. For the two-electron state Co-PSB is lifted for either L-DQD or R-DQD by the spin exchange between the left or right dot and the adjacent reservoir (source) just the same mechanism shown in (f). On the other hand, for the three-electron state Co-PSB is lifted in the same way but independently for the two DQDs, and in addition, Co-PSB lifting in either L- or R-DQD can subsequently lift Co-PSB in the other DQD because the spin in the center dot can be flipped in the first Co-PSB lifting process. Therefore Co-PSB is lifted efficiently [22] at the (0,1,1)-(1,1,0) degenerate point as shown in Fig. 3(g). As a result the total transport current due to Co-PSB lifting can be increased at the degenerate point (shown by the green circle) from just the sum of the current measured at the green triangle and square in Fig. 3(a). Although less distinct, the same effect is also observed for the weak inter-dot tunnel coupling even when current along the charging lines is much less pronounced (data is not shown).

Next we move on to the unexpected CB lifting observed as sharp lines in the CB region of (0,1,1), and (1,1,0) marked by the yellow square, and circle, respectively in Fig. 3(b). These lines are not observed under $B_{ext} = 0$ T (see Fig. 3(a)). Here, the lowest excited state is $(0,2,0)S_C$ and co-tunneling through $(0,2,0)S_C$ is spin blocked (Co-PSB) once a triplet state is formed in L-DQD or R-DQD. These CB lifting lines can be explained by resonace between two charge states, both of which are the excited states. We attribute the two excited states to $(1,1,0)T_{+,LC}$ and $(0,2,0)S_C$ at the yellow square point, and $(0,1,1)T_{+,CR}$ and $(0,2,0)S_C$ at the yellow circle point, respectively (see Fig. 3(h) and (i)) since these CB lifting lines are almost aligned with the transitions for the three-electron states between $(0,2,1)S_C$ and $(1,1,1)S_{LC}$ and between $(1,2,0)S_C$ and $(1,1,1)S_{CR}$.

On resonace of $(0,2,0)S_C$ with $(1,1,0)T_{+,LC}$, or $(0,1,1)T_{+,CR}$, a S-T hybridized state is formed mediated by nuclear field fluctuations, which provides a co-tunneling channel to lift Co-PSB of R-DQD, or L-DQD, respectively (see Fig. 3(h) and (i)). This interpretation is supported by calculation of the electro-chemial potentials for the two-electron ground and excited states in Fig. 3(j). The figure plots the electro-chemical potentials of the two electron states as a function of energy detuning between the leftmost and rightmost dot (the detuning is defined on the stability diagram, see Fig. 3(b)). Here, $E_t$ of a few tens of μeV ($\gg \Delta E_n$) is assumed. The cross points of two lines between $(1,1,0)T_{+,LC}$ and $(0,2,0)S_C$, and between $(0,1,1)T_{+,CR}$ and $(0,2,0)S_C$, marked by the black circles in Fig. 3(j), correspond to the yellow square, and the yellow circle, respectively in Fig. 3(b). Here Co-PSB lifts due to the formation of a S-T hybridized state. This S-T hybridization only occurs around $B_{ext} \sim 1$ T where Zeeman energy compensates $E_t$ ($\Delta E_{ST} \lesssim \Delta E_n$), similar to the case of PSB lifting of DQDs having a strong inter-dot tunnel coupling (Fig. 2(g)). The Co-PSB lifting line is not clear on the positive detuning side. This is probably because the line position, which is almost on top of the base line of the bias triangle, is very close to the zero detuning position where the electro-chemical potentials of (1,1,0) and (0,1,1) cross as shown by the green circle in Fig. 3(a).

To confirm the Co-PSB lifting observed in Fig. 3(a) and (b), we tune the center plunger gate voltage to energetically lower the (0,2,0) state (either $(0,2,0)S_C$ or $(0,2,0)T_\alpha$) such that the (0,1,1)-(1,1,0) transition is replaced by the (0,1,1)-(0,2,0)-(1,1,0) transition (see Fig. 4(a) and (b)). Then, transport through the (0,2,0) state is Coulomb blockaded and no Co-PSB lifting is expected since there is no excited state available for co-tunneling. Indeed, we observe no excess current along the (0,1,1)-(0,2,0)-(1,1,0) tranition (white circles) in the biased stability diagram of Fig. 4(a) and (b) while a finite co-tunneling current is observed at the green circle in Fig. 3(a) and (b).

Finally we discuss an additional unexpected feature of CB lifting observed as a sharp line only under a finite external magnetic field in the (1,1,1) CB region located between two bias triangles (see the black circle in Fig. 3(b)). This line can be also accounted for by Co-PSB lifting whose mechanism is similar to that observed in the (0,1,1) CB region (see Fig. 3(h)). The difference is that now the co-tunneling is mediated by the coupling between different excited states. We attribute the two excited states to $(1,1,1_P)$ and $(0,2,1)S_C$, where the subscript P indicates that the electron occupies the first excited orbital state (see Fig. 5(a) and (b)). Fig. 5(c) shows the calculated energy diagram of three electron states, (1,2,0), (1,1,1) and (0,2,1). The two black circles indicate the cross points between $(0,2,1)S_C$ and $(1,1,1)T_{+,LC}$. The upper (lower) circle corresponds to the case of $(0,2,1)S_C$ with the rightmost dot having a down (up) spin electron. Here, these states are mixed by the inhomogeneous hyperfine field, and as a consequence, PSB is lifted at the black square in Fig. 3(b), as usually observed in DQDs. On the other hand, the two grey circles in Fig. 5(c) show the cross points between $(0,2,1)S_C$ and $(1,1,1_P)$ where these states are also mixed by the inhomogeneous hyperfine field to form an intermediate state which is available for co-tunneling, lifting Co-PSB to generate a current line in the (1,1,1) CB region (black circle in Fig. 3(b)).

**Conclusion:**

In conclusion, we design and fabricate a 3T-TQD device to study PSB in the strongly tunnel-coupled three dots. We observe PSB in both of L-DQD and R-DQD in the biased stability diagram as often observed for double dots but with distinct influences from the third dot. We investigate the effect of PSB to suppress not only the first order tunneling current but also the co-tunneling current for the first time, and find that this co-tunneling PSB is lifted due to resonant coupling of excited states to the third dot and also spin exchange of the two side dots with the adjacent leads. The present result indicates that PSB is also important in the second order process, in addition to the first order process as usually discussed.

**Acknowledgement:**

Part of this work is financially supported by the ImPACT Program of Council for Science, Technology and Innovation (Cabinet Office, Government of Japan) the Grant-in-Aid for Scientific Research (No. 26220710), CREST (JPMJCR15N2, JPMJCR1675), JST, Incentive Research Project from RIKEN. AN acknowledges support from Advanced Leading Graduate Course for Photon Science (ALPS). TT and JY acknowledges support from JSPS Research Fellowships for Young Scientists. TN acknowledges financial support from JSPS KAKENHI Grant Number 25790006. TO acknowledges financial support from Grants-in-Aid for Scientific Research (No. 16H00817, 17H05187), PRESTO (JPMJPR16N3), JST, Yazaki Memorial Foundation for Science and Technology Research Grant, Advanced Technology Institute Research Grant, the Murata Science Foundation Research Grant, Izumi Science and Technology Foundation Research Grant, TEPCO Memorial Foundation Research Grant, The Thermal & Electric Energy Technology Foundation Research Grant, The Telecommunications Advancement Foundation Research Grant, Futaba Electronics Memorial Foundation Research Grant, MST Foundation Research Grant.

**Figures:**

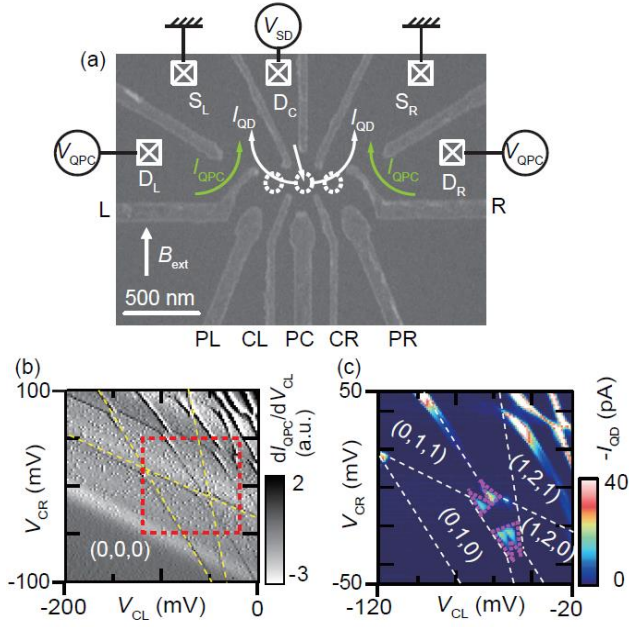

Fig. 1. (a) Scanning electron micrograph of the 3T-TQD device identical to the one measured. The white dashed circles indicate the positions of the three dots. Ohmic contacts used for the source ($S_L$, $S_R$) and drain ($D_C$, $D_L$, $D_R$) electrodes are shown by the white boxes. The white arrows show the current paths through the three dots. We apply $V_{SD}$ to $D_C$ and ground other Ohmic contacts ($V_{QPC} = 0$ mV) to measure the current flowing from $D_C$ to the ground ($S_L$ and $S_R$) as $I_{QD}$. The green arrow shows the current path of the QPC charge sensor used for the experiment. We apply $V_{QPC} = 1$mV to $D_L$ and $D_R$ while $D_C$, $S_L$, and $S_R$ are grounded to measure $I_{QPC}$. Here, no bias is applied across both of the DQDs ($V_{SD} = 0$ mV). (b) Charge stability diagram as a function of gate voltages $V_{CL}$ and $V_{CR}$ obtained by charge sensing. The yellow dashed lines with different slopes indicate the charging lines for the respective dots. The steepest yellow dashed line for the faint charging line of the left dot is drawn by smoothly connecting the anti-crossing features of the charging lines observed due to the coupling between the left and the center dots at ($V_{CL}$, $V_{CR}$) = (-45 mV, -25 mV) and (-70 mV, 75 mV). (c) Charge stability diagram biased with $V_{SD}$ = -0.5 mV obtained by transport measurement. The gate voltage condition is the same as in (b) but only we measure in a smaller region of $V_{CL}$ and $V_{CR}$ shown by the red dashed box in (b). The purple dashed lines are the guides for the bias triangles formed near the boundaries of (0,1,1)-(0,2,0) and (0,2,0)-(1,1,0).

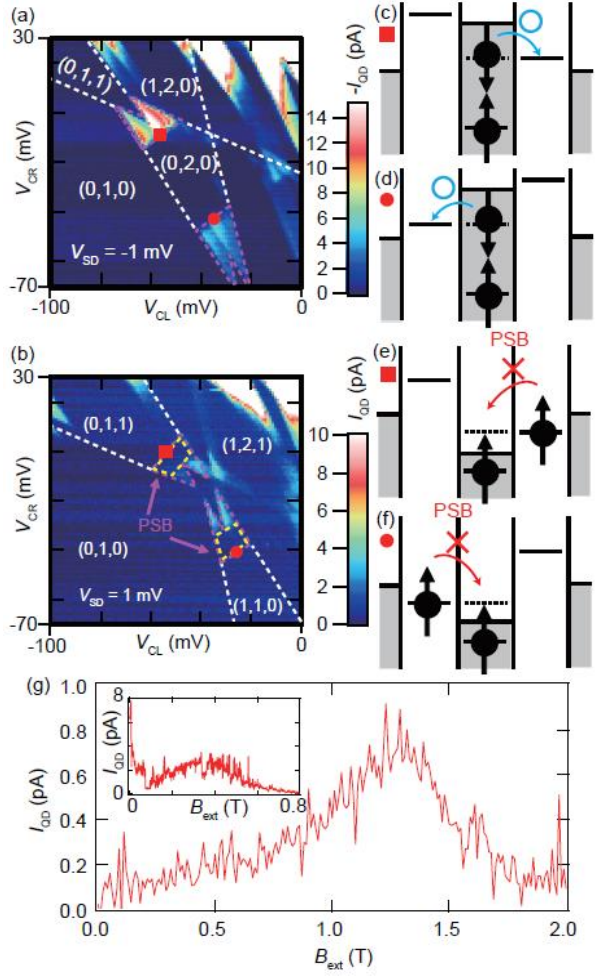

Fig. 2. (a), (b) Biased stability diagram obtained by the transport measurement of $I_{QD}$ as a function of $V_{CL}$ and $V_{CR}$ at $B_{ext}$ = 0 T with $V_{SD}$ = -1 mV (a) and $V_{SD}$ = 1 mV (b). PSB is only observed for the positive $V_{SD}$ in (b) as the current suppression inside the yellow dashed trapezoidal regions in the bias triangles. (c) - (f) Energy level diagrams on the resonance of $(0,1,1)S_{CR}$-$(0,2,0)S_C$, and $(0,2,0)S_C$-$(1,1,0)S_{LC}$ at the red square, and circle in (a), and (b), respectively. Each of colored arrow shows an electron tunneling path. Grey regions represent the Fermi distribution in the respective lead. (g) Leakage current at the singlet resonance line of L-DQD (red circle in (b)) as a function of $B_{ext}$. We note that R-DQD also shows a similar $B_{ext}$ dependence indicating that the inter-dot coupling is roughly symmetric. Inset: same measurement performed under the weak inter-dot coupling condition.

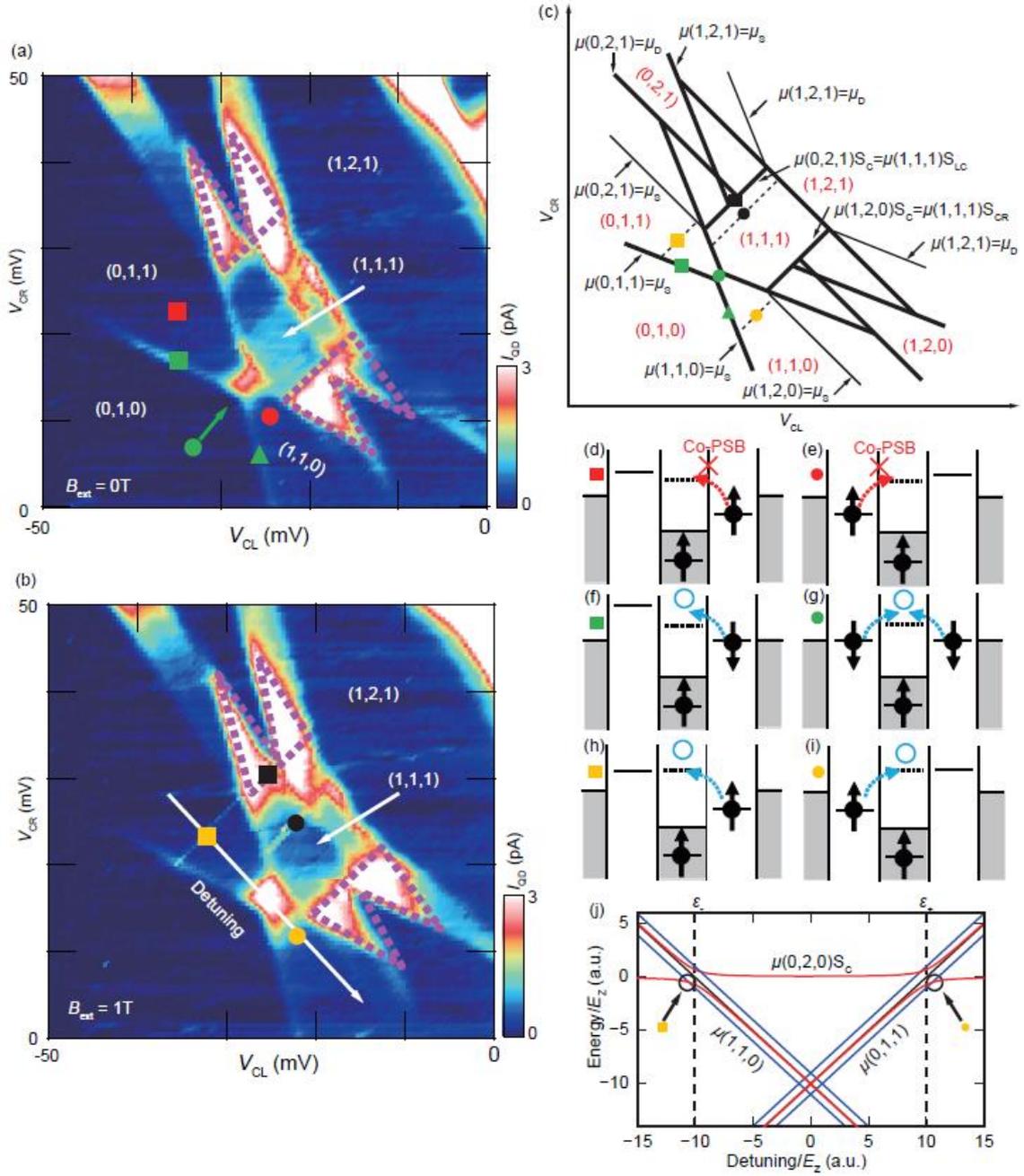

Fig. 3. (a), (b) Stability diagram with $V_{SD} = 1$ mV measured for $B_{ext} = 0$ T (a) and $B_{ext} = 1$ T (b). The purple dashed lines are the guides of the bias traingles. (c) Schematic of a stability diagram under positive $V_{SD}$ which corresponds to the situation of (a) and (b). At the (0,1,1)-(1,1,0) boundary marked by the green circle, $\mu(0,2,0)S_C$ is energetically higher than $\mu(0,1,1)$ and $\mu(1,1,0)$. (d) - (i) Energy level diagrams in the (0,1,1) and (1,1,0) CB regions as indicated by the red, green, and yellow squares and circles in (a) and (b), respectively. The dashed arrows show the co-tunneling path through $(0,2,0)S_C$ as the intermediate state. Co-PSB is observed when $(1,1,0)T_{a,LC}$ or $(0,1,1)T_{a,CR}$ is formed ((d), (e)). Co-PSB can be lifted due to the exchange of electrons between the side QD and the adjacent lead ((f),

(g)). At $B_{ext}$ = 1 T, Co-PSB can also be lifted due to resonant coupling of excited states to the third dot level ((h), (i)). (j) Energy diagram of two-electron states as a function of energy detuning between the leftmost and rightmost dots. The ground state is (1,1,0) and (0,1,1) for the positive and negative detuning, respectively. The zero-detuning is at the cross point of $\mu(0,1,1)$ and $\mu(1,1,0)$ marked by the green circle in (a). Two dashed black lines indicate the positions of singlet resonances between (1,1,0)$S_{LC}$-(0,2,0)$S_C$ at detuning $\varepsilon_-$ and (0,1,1)$S_{CR}$-(0,2,0)$S_C$ at $\varepsilon_+$ (see Fig. 3(j)). The red, black, and blue lines show singlet, $T_0$, and $T_+$ or $T_-$ states, respectively with positive, zero and negative slopes for (0,1,1), (0,2,0) and (1,1,0) charge configurations. The energy scale is normalized by $E_Z$. For the calculation we use a symmetric inter-dot coupling $E_t = E_Z$ and $\varepsilon_\pm = \pm 10 E_Z$ (the cross point of $\mu(0,1,1)$ and $\mu(0,2,0)S_C$ for $\varepsilon_+$ and that of $\mu(1,1,0)$ and $\mu(0,2,0)S_C$ for $\varepsilon_-$) which roughly agrees with the experiment.

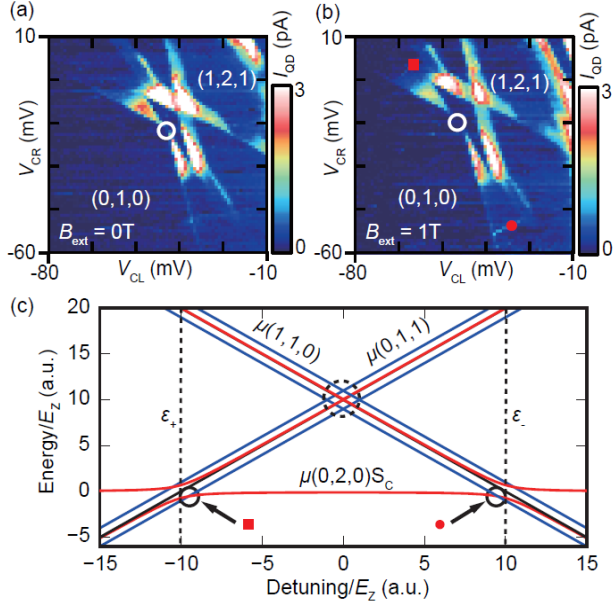

Fig. 4. (a), (b) Stability diagrams measured for a strongly coupled 3T-TQD with $V_{SD} = 1$ mV at $B_{ext} = 0$ T (a) and 1 T (b). The difference from Fig. 3(a) and (b) is the energy level of the center dot such that $\mu(0,2,0)S_C$ is lower than the resonance of $\mu(1,1,0) = \mu(0,1,1)$. The white open circles show the positions that satisfy the condition of $\mu(1,1,0) = \mu(0,1,1) = \mu_S$. PSB lifting is observed around the red square and circle in (b) as usually observed in DQDs. (c) Energy diagram of two-electron states as a function of $\varepsilon$ calculated using the same parameter for Fig. 3(j) except for $\varepsilon_{\pm} = \mp 10E_Z$. The black dashed circle shows the point where $\mu(1,1,0) = \mu(0,1,1) = \mu_S$ (white circle in (b)).

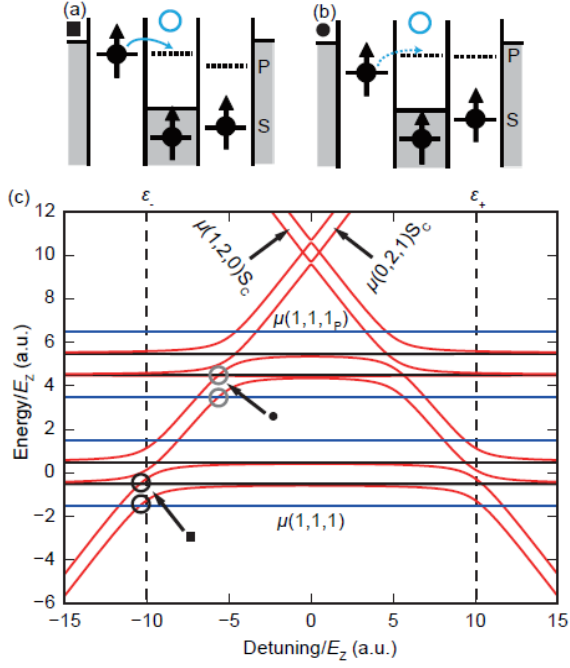

Fig. 5. (a), (b) Energy level diagrams that correspond to the black square (a) and circle (b) in Fig. 3(b) where an electron tunnels from the left QD to the center QD due to PSB (a) and Co-PSB (b) lifting. The blue arrow in (a) shows an electron tunneling path from (1,1,1) to (0,2,1)$S_C$ while the blue dashed arrow in (b) shows an electron co-tunneling path through (0,2,1)$S_C$ as the intermediated state. (c) Energy diagram of three electron states as a function of $\varepsilon$. The red lines with positive and negative slopes show the Zeeman split states of (0,2,1)$S_C$ and (1,2,0)$S_C$, respectively, while the lower and upper four horizontal lines show those of (1,1,1) and (1,1,1$_P$), respectively (having different $S_z$ component, $\pm 3/2$ and $\pm 1/2$). The energy scale is normalized by $E_Z$. We assume a symmetric inter-dot coupling $E_t$ = $E_Z$, $\mu(1,1,1_P)$ - $\mu(1,1,1)$ = $5E_Z$ and $\varepsilon_\pm$ = $\pm 10 E_Z$. $\mu(1,2,0)S_C$ and $\mu(1,1,1)S_{CR}$ cross at $\varepsilon_+$ and $\mu(0,2,1)S_C$ and $\mu(1,1,1)S_{LC}$ at $\varepsilon_-$. $\mu(1,2,0)S_C$ and $\mu(0,2,1)S_C$ cross at zero detuning.